\documentclass[usenatbib,letter]{mn2e}
\usepackage{psfig}
\usepackage[paperwidth=8.5 in,paperheight=11in,left=2.0cm,right=2.0cm,top=2cm]{geometry}

\def\mnras{{MNRAS}}
\def\apj{{ApJ}}
\def\aj{{AJ}}
\def\aap{{A\&A}}
\def\apjl{{ApJL}}
\def\apjs{{ApJS}}
\def\araa{{ARAA}}
\def\nat{{Nature}}
\def\apss{{ApSS}}

\def\xmm{{\sl XMM-Newton}}
\def\asca{{\sl ASCA}}
\def\xte{{\sl RXTE}}

\def\ark{{Ark~564}}
\def\mcg6{{MCG-6-30-15}}

\def\mch{M$\rm^{c}$Hardy\,}

\def\ltsim{\mathrel{\hbox{\rlap{\hbox{\lower4pt\hbox{$\sim$}}}\hbox{$<$}}}}
\def\gtsim{\mathrel{\hbox{\rlap{\hbox{\lower4pt\hbox{$\sim$}}}\hbox{$>$}}}}

\title[Multiple Lorentzians in Ark 564]{ Discovery of multiple Lorentzian
  components in the X-ray timing properties of the Narrow Line Seyfert
  1 Ark 564}

\author[I. M. McHardy et al.]
{I. M. McHardy$^{1}$\thanks{E-mail: imh@astro.soton.ac.uk},
P. Ar\'evalo$^{1}$, P. Uttley$^{1}$,  I. E. Papadakis$^{2}$, D. P.
Summons$^{1}$,
\and W. Brinkmann$^{3}$, M. J. Page$^4$\\
$^1$School of Physics and Astronomy, University of Southampton,
Southampton S017 1BJ, UK \\
$^2$Physics Department, University of Crete, P.O.Box 2208, 71003
Heraklion, Greece\\
$^3$ Max-Planck-Institut f\"ur extraterrestrische Physik,
Giessenbachstrasse, 85741 Garching, Germany\\
$^4$ Mullard Space Science Laboratory, University College London,
Holmbury St Mary, Dorking, RH5 6NT\\
}

\begin{document}
\date{Received /Accepted}
\pagerange{\pageref{firstpage}--\pageref{lastpage}} \pubyear{2006}

\maketitle
\label{firstpage}
\begin{abstract}
We present a power spectral analysis of a 100~ksec \xmm\ observation of
the narrow line Seyfert 1 galaxy Ark~564. When combined with
earlier \xte\ and \asca\ observations, these data produce a
power spectrum covering seven decades of frequency which is well
described by a power law with two very clear breaks. This shape is
unlike the power spectra of almost all other AGN observed so far, which
have only one detected break, and resemble Galactic binary systems in
a soft state.  The power spectrum can also be well described by the sum
of two Lorentzian-shaped components, the one at higher frequencies
having a hard spectrum, similar to those seen in Galactic binary
systems. Previously we have demonstrated that the lag of the hard band
variations relative to the soft band in Ark~564 is dependent on
variability time-scale, as seen in Galactic binary sources.  Here we
show that the time-scale dependence of the lags can be described well
using the same two-Lorentzian model which describes the power spectrum,
assuming that each Lorentzian component has a distinct time lag.  Thus
all X-ray timing evidence points strongly to two discrete, localised,
regions as the origin of most of the variability. Similar behaviour is
seen in Galactic X-ray binary systems in most states other than the
soft state, i.e. in the low-hard and intermediate/very high states.  Given the
very high accretion rate of Ark~564 the closest analogy is with the
very high (intermediate) state rather than the low-hard state.  We
therefore strengthen the comparison between AGN and Galactic binary
sources beyond previous studies by extending it to the previously
poorly studied very high accretion rate regime.

\end{abstract}

\section{Introduction}

Understanding the relationship between AGN, which are powered by
accretion onto supermassive black holes, and the stellar mass Galactic
black hole binary systems (GBHs) is currently one of the major
research areas in high energy astrophysics. If we can
understand the relationship, then we can predict how AGN should behave
on cosmological time-scales by observing the brighter, and much faster
varying, GBHs.

The comparison is complicated by the fact that (at the risk of some
oversimplification), GBHs can exist in a number of states, defined
originally in terms of their medium energy (2--10 keV) X-ray properties,
particularly: (i) the hard spectrum, low-flux (`hard') state, (ii) the
soft spectrum, high flux (`soft') state and (iii) the `very high
state' (VHS - sometimes also described as the high-intermediate state)
where the flux is very high and the spectrum very
soft. In general the accretion rate increases as we go from the hard
to soft to VHS states. See \cite{remillard06} for a comprehensive
description of GBH states and behaviour.

The various states have quite distinct X-ray timing properties which
may, in fact, provide a more fundamental state discrimator than the
spectral properties. Timing is usually quantified in terms of the
power spectral densities (PSDs) of the X-ray lightcurves. Where
significant variability is observed, `soft' state PSDs are
characterised by a power law of slope of -1 at low frequencies
(i.e. the power, $P(\nu) \propto \nu^{-\alpha}$ with $\alpha\sim 1$),
breaking to a slope $\alpha \geq 2$ above a characteristic frequency
$\nu_{B}$ (or time-scale $T_{B}$). Hard state and VHS state GBH PSDs,
which are similar to each other, are more complex. They are best
described by the combination of two or more Lorentzian-shaped
components \citep[e.g.][]{nowak00}, although at low signal to noise
they can be approximated by a power law with a second, lower frequency
break, approximately 1.5 to 2 decades lower in frequency than the high
frequency break.

So far, almost all well observed AGN, which are X-ray bright and have
moderately high accretion rates, have soft state PSDs, as we would
expect from comparison with GBHs of similar accretion rate
(measured relative to the Eddington accretion rate, $\dot{m}_{E}$)
\citep[e.g.][]{mch04,mch05,uttleymch05}. However the comparison for
the hard and VHS states is largely unknown. Until very recently,
NGC3783 had been suggested to have two breaks in its PSD
\citep{markowitz03a} and so be a hard state AGN, although its
accretion rate is similar to that of the well observed soft state
AGN. However recent work, based on improved data \citep{summons07},
has shown that, in fact, NGC3783 is another soft state AGN.
If soft states are linked to the accretion disc extending in to very
small radii, then perhaps their much cooler accretion discs compared to
GBHs may allow the optically thick disc to survive without evaporation to
smaller radii. Thus maybe all AGN are soft state objects and the
comparison with GBHs does not extend to other states?
Such a conclusion would have a
fundamental impact on our understanding of the accretion process onto
black holes and so it is very important to confirm, from timing
properties, a definite non-soft state AGN.

The only remaining case where there is at least some evidence for a
non-soft state PSD is \ark.  From observations with \xte, evidence
for a low frequency break in the PSD
was presented by \cite{pounds01}. Evidence for a second break, at high
frequencies, was presented by \cite{papadakis02} from analysis of
\asca\ observations. \cite{papadakis02} interpreted the combined
\xte\ and \asca\ PSD as a hard state PSD but did note that the high
frequency break does not scale linearly with mass to the hard state of
the archtypal GBH Cyg X-1; the frequency difference is too small.
However as \ark\ is one of the highest accretion rate AGN
known ($\dot{m}_{E}\sim 1$, \cite{romano04}), a VHS state remains a
strong possibility.

Unfortunately proper modelling of the combined long and short
time-scale PSD is hampered by the gaps which occur in the
\asca\ light curves, and hence in the PSD, at the orbital period ($\sim
5600$s). We therefore made a 100 ks continuous observation with
\xmm\ to fill that gap in the PSD. Spectral analysis of this
\xmm\ observation has been presented by \cite{papadakis07} and
\cite{brinkmann07} and cross-spectral analysis (i.e. lags between
energy bands as a function of Fourier frequency or time-scale) has been
presented by \cite{arevalo06}.

Here we present a full power spectral analysis of the combined \xte,
\asca\ and \xmm\ data and find firm evidence for two breaks in the
combined PSD. We also show that the PSD can be parameterised as the
sum of two Lorentzian components. We note particularly that the
highest frequency Lorentzian, for which good spectral information is
available from both the \xmm\ and \asca\ observations, is stronger at
higher energies, exactly as is found for Lorentzian components in
GBHs. In GBHs in the hard and VHS states, changes in the lag between
bands occur at approximately the time-scales at which the PSD changes
from one Lorentzian component to another, supporting the model that
each Lorentzian comes from a separate physical emission region.  In
this paper we therefore re-examine the lag spectrum and compare it
with the PSD, to test whether a two-component emission region may
apply in \ark\ and also as a further diagnostic of the state of
the AGN.

The paper is arranged as follows: In Sec. \ref{data}, we describe the
observational data
sets which we used and discuss the individual
\asca\ and \xmm\ PSDs in Secs~\ref{s_ascapsd} and \ref{s_xmmpsd}. We
then discuss the joint \xte\, \asca\ and \xmm\ power spectrum and its
energy dependence in Sec.~\ref{s_combinedpsd} and compare the powerspectrum
with the lag spectrum in Sec~\ref{s_lags}. We summarise our conclusions
in Sec.~\ref{s_conclusions}.

\section{Summary of observations}
\label{data}
\subsection{\xte}

\ark\ was observed by \xte\ from 1999 January 1 to 2003 March 4. For
most of this period observations were carried out approximately once
every 4 days but from 2000 May 31 to 2000 July 01, observations were
carried out 8 times per day. The observations from 1999 January 1 to
2000 September 19 are described in detail in \cite{pounds01}. We have
extracted all of the \xte\ data from the archive and produced 2--10
keV lightcurves in the same manner as we have described for other
\xte\ AGN monitoring programmes \citep[e.g. see][for details]{mch04}.
For use in power spectral analysis we produce a long time-scale light
curve, covering the full observing period with 4-day sampling, and an
intermediate time-scale lightcurve, covering the one month period of
3-hour sampling.

\subsection{\asca}

\ark\ was monitored with \asca\ between 2000 June 1 and July 5,
producing a month-long light curve. This observation was only
interrupted by the regular Earth occultations of the satellite on
the orbital time-scale of 5632s. Binning the data in orbit-long
bins produces an evenly sampled light curve, containing 551
points, that probes time-scales from a few times $10^6$ to $10^4$
s. Additionally, each of the 551 individual orbits cover
time-scales of $10^3-10^2$ s, whose power spectra can be combined
to produce a high quality high frequency PSD.

We downloaded SIS data from the TARTARUS database in 0.6--2 and 2--10
keV energy bands and produced background subtracted light curves, with
average count rates of 2.7 and 0.6 c/s respectively. The average
photon energies of these two bands are 1.18 and 3.2 keV respectively.

The lightcurves,
PSD and other variability properties of this data set have been
studied by e.g. \citet{papadakis07,edelson02}.

\subsection{\xmm }

\ark\ was observed by \xmm\ from 2005 January 5, 19:47, to 2005 January
6, 23:16. The European Photon Imaging Cameras (EPIC) PN, MOS1 and MOS2
cameras were all operated in small window mode using a medium filter,
as described by \cite{papadakis07}. The total exposures for the PN
and MOS cameras were 98.8 ks and 99.1 ks respectively. For the PN
camera, we extracted source photons from a $\sim 2 \arcmin \times 2
\arcmin$ region and the background was selected from a source-free
region of equal area on the same chip. Only single and double events,
with quality flag=0 were retained. For the MOS1 and MOS2 cameras,
source photons were extracted from a circular region of $\sim 46
\arcsec$ in radius and single, double, triple and quadruple events
were used. The PN data was free of pile-up, but both MOS cameras
suffered significant pile-up in the cores of the target
PSFs. Therefore we discarded data from the central $12\arcsec$
diameter of each MOS exposure. The average PN background-subtracted
count rate in the 0.2--10 keV band was $\sim 28$ c/s.  The background
level was low and almost constant throughout the observation.

\begin{figure}
\psfig{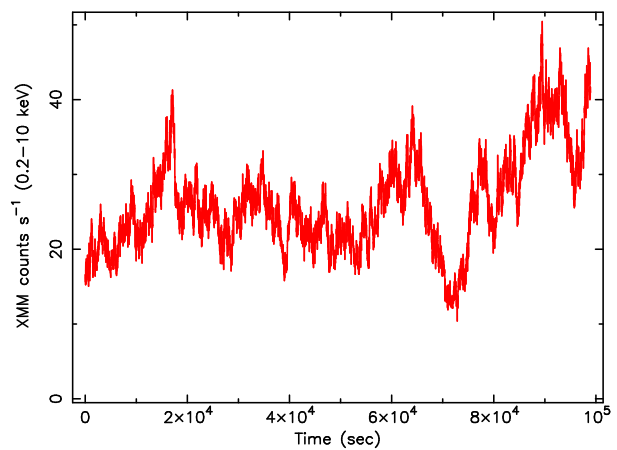}
\caption{\label{fig:xmmlc} A weighted average combined PN, MOS1 and
MOS2 \xmm\ lightcurve in the 0.2--10~keV band.}
\end{figure}

We produced light curves in the 0.2--2 and 2--10 keV energy bands in
48 s bins, and combined the light curves from the different detectors
to produce weighted average light curves. The 0.2--10~keV
lightcurve is shown in Fig~\ref{fig:xmmlc}  \citep[see also][]{papadakis07}.
As the variability properties normally depend on energy band, we also
constructed \xmm\ light curves from the PN data alone in the 0.5--2
and 2--8.8 keV bands, to match the average energies of the \asca\ light
curves.  We used these lightcurves in all quantitative analyses.

\section{\asca\ Power spectrum}
\label{s_ascapsd}

\begin{figure}
\psfig{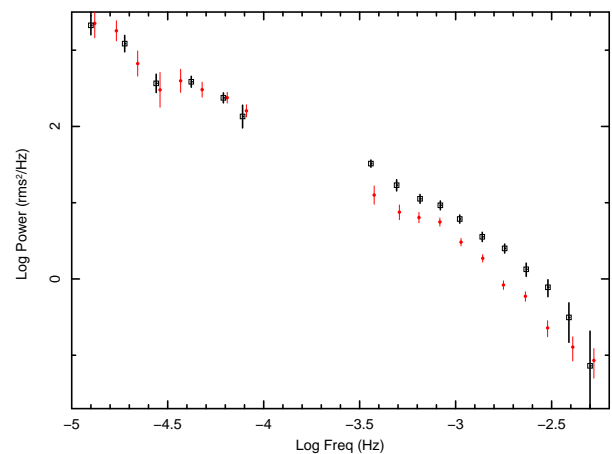}
\caption{\label{fig:ascapsds}\asca\ PSDs. The open (black) squares,
  with thick lines, are the 2--10 keV data. The filled (red) circles,
  with thin lines, are the 0.6--2 keV data.  The Poisson noise level
  has been removed from both datasets.  Data below $10^{-4}$Hz result
  from lightcurves binned on the orbital time-scale.  Data above
  $10^{-3.5}$Hz are the sum of the PSDs derived from the many short
  sub-orbital lightcurves. The 2--10 and 0.6--2 keV data points have
  been shifted very slightly in frequency to avoid overlap at both
  ends of the overall PSDs.  }
\end{figure}

The superposed PSDs derived from the \asca\ 2--10 keV and 0.6--2 keV
data are shown in Fig.~\ref{fig:ascapsds}. Data below $10^{-4}$Hz
result from the light curves binned on the orbital time-scale.  Data
above $10^{-3.5}$Hz are the sum of the PSDs derived from the many
short sub-orbital light curves.  Note the frequency gap
that these data do not cover.
The Poisson noise level has been
subtracted.  On time-scales longer than the orbital time-scale, the low
and high energy PSDs have a similar normalisation. However on shorter
time-scales the 2--10 keV PSD lies above the 0.6--2 keV PSD, except perhaps
at the very highest frequencies ($\sim 10^{-2}$Hz) where the two PSDs
may be converging. This frequency dependent excess of the high energy PSD
over the low energy PSD both below and above the high frequency break ($\sim
10^{-3}$Hz), which is different from the behaviour of other well
observed AGN, \citep[e.g. NGC4051,][]{mch04}, will be modelled properly in
Section~\ref{s_combinedpsd}.

\section{\xmm\ Power spectrum}
\label{s_xmmpsd}

\begin{figure}
\psfig{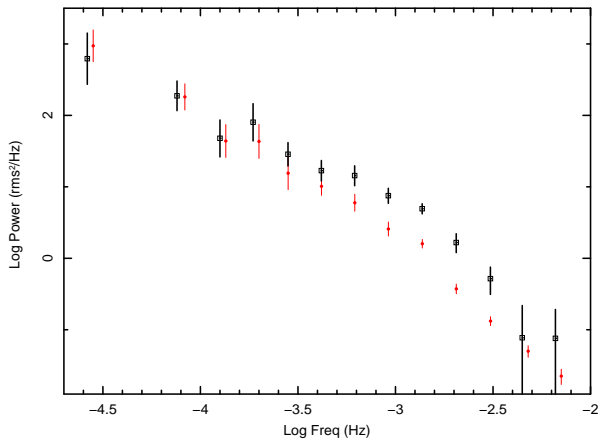}
\caption{\label{fig:xmm210psds}\xmm\ PSDs. Here, for display purposes,
  we show the 2--10 keV (upper data points, black open squares, thick
  lines) and 0.2--2 keV (lower, red filled circles, thin lines) PSDs,
  with Poisson noise level subtracted, as
  the band centroids are more separated than for the 0.5--2 and 2--8.8
  keV bands. Thus spectral differences, particularly the excess around
  $10^{-3}$Hz at higher energies, are a little more pronounced. We do,
  however, use the 0.5-2 and 2-8.8 keV bands (Table~\ref{tab:brpl})
  for quantitative analyses, to match better the \asca\ energy bands.}
\end{figure}

The \xmm\ light curves were almost completely continuous except for 5
small gaps of duration 20-30 secs each. Following the procedure
established in \cite{mch04} these gaps were filled by linear
interpolation, together with the addition of noise characteristic of
the surrounding data points, in order to eliminate spurious high
frequency power in the power spectrum. Power spectral densities
were derived from these light curves using a standard direct Fourier
transform.  In Fig. ~\ref{fig:xmm210psds} we show the 0.2--2 and 2--10 keV
PSDs and we see close agreement with the \asca\ PSDs in the same
energy bands. As with the \asca\ PSDs, the low and high energy
\xmm\ PSDs coincide very closely at frequencies below $10^{-4}$Hz, but
at higher frequencies the 2--10 keV PSD exceeds the lower energy
PSD. Note that here, for the low energy band, we show the 0.2--2 keV
PSD rather than the 0.5--2 keV PSD and note that the bump centred
around $10^{-3}$Hz is even
less pronounced in the 0.2--2 keV band, strengthening our observation
that the excess has a strongly energy-dependent shape.  Although not
shown, the \xmm\ 0.5--2 keV PSD is almost identical to the \asca\ 0.6--2
keV PSD.

A simple power law fit to both low and high energy PSDs is a very poor
fit, leaving a large residual bump centred around $10^{-3}$Hz. A
bending powerlaw, however, which provides a good fit to the high/soft
state
of Cyg X-1 and to the \xmm\ PSDs of NGC4051 \citep{mch04} and \mcg6
\citep{mch05}, is a good fit (Table~\ref{tab:brpl}).  Tying the break
frequencies to be the same in both energy band PSDs, we note (as can
also be seen
directly from Fig.~\ref{fig:xmm210psds}), that although the slopes
above the break are similar in both energy bands, the slope below the
break is noticeably flatter at higher energies.  The \asca\ PSDs,
although less well defined because of the gap on the orbital
time-scales, are quite consistent with the fit parameters given in
Table~\ref{tab:brpl}.  This behaviour is different from that of
NGC4051 and MCG-6-30-15 \citep[e.g. see Table 6 of][]{mch05} where,
although the slopes below the break are not very well determined,
there is no evidence for a variation with energy. Above the
break, however, the slope is flatter at higher energies in NGC4051 and
MCG-6-30-15.

The PSD slope variations seen here in \ark\ can be more naturally
interpreted in terms of a power law (which is necessary to explain the
power at low frequencies) together with a Lorentzian component which
is stronger at higher energies.  Such a model, in which the power law
slope and normalisation, and Lorentzian central frequency are tied,
provides an equally good fit to the data.  In the case of the latter
model, we note that the Lorentzian component is approximately three
times stronger in the 2-8.8 keV than 0.5--2 keV band
(Table~\ref{tab:lor}). An increasing Lorentzian amplitude at higher
energies is also typically observed in QPOs and some broad Lorentzian
components in GBHs \citep[e.g.][]{pott03}.  As the exact values
of the fit parameters depend on proper fitting of the lower frequency
data from \xte\ and \asca\, we leave further discussion to
Section~\ref{s_combinedpsd}.

\begin{figure}
\psfig{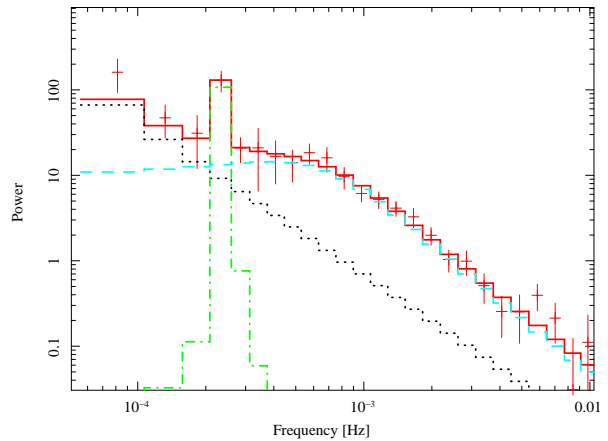}
\caption{\label{fig:xmm2lor} \xmm\ 2--8.8 keV \xmm\ PSD showing the model
  components. The broad high frequency Lorentzian is shown by a dash
(light blue) line, the narrow Lorentzian is shown by a dot-dash (green) line,
  the underlying power law component is a dotted (black) line, the
  data points are (red) crosses, and the overall model is a solid
  (red) line.  The Poisson noise level has been subtracted. }
\end{figure}

There is a residual to the power law plus Lorentzian fit to the \xmm\
PSD at $\sim 2.2 \times
10^{-4}$Hz. This residual can be fitted (using the unbinned data) by
the addition of a second, narrow (width $\sim8 \times
10^{-6}$Hz), Lorentzian with normalisation approximately half that of
the broad Lorentzian.  The fit is shown in Fig.~\ref{fig:xmm2lor}.  The
fit is improved but, adjusting the F-test probabilities to take
account of the fact that the line could have appeared in any of the
available frequency bins, we find that the improvement is significant
only at the 90\% confidence level.

\begin{table*}
\centering
\caption {\bf Bending Powerlaw: combined fit to the \xmm\ soft and
  hard band data}
\begin{tabular}{crrccc}
Energy Band & $\alpha_{mid}$ & $\alpha_{high}$ & $\nu_{B}$ & red
$\chi^{2}$ & d.o.f\\
keV  &  &  & Hz & & \\
\hline
0.5--2 & $1.31\pm 0.12$ & $3.62\pm0.37 $ & $1.73\pm0.44\times 10^{-3}$ & &  \\
2.0--8.8 & $0.95\pm0.14$ & $3.57\pm0.61$ & tied & 0.88 & 37 \\
\end{tabular}
\label{tab:brpl}
\end{table*}

\begin{table*}
\centering
\caption {\bf Power Law with Lorentzian Line: combined fit to the
  \xmm\ soft and hard band data}
\begin{tabular}{clllccc}
Energy Band & $\alpha$ &  Line centre  & Line width & Line
normalisation & red $\chi^{2}$ & d.o.f\\
keV  &  &   Hz & & & & \\
\hline
0.5--2.0 & $1.80\pm0.19$ & $3.94\pm1.7\times 10^{-4}$ &$6.8\pm16\times
10^{-4} $ &$8.8\pm5.7 \times 10^{-3}$  &  & \\
2.0--8.8 & tied & tied & $1.1\pm0.17\times10^{-3}$ & $24.1\pm8.4
\times 10^{-3}$ &1.05 & 35 \\
\end{tabular}
\label{tab:lor}
\end{table*}

\section{Combined \xmm, \asca\ and \xte\ Power spectra}
\label{s_combinedpsd}

The \xte\ data, which covers time-scales from $\sim3$ years to 3 hours,
has already been analyzed by \cite{pounds01} and, including further
data, by \cite{markowitz03a} who both fit a breaking power law model
to the PSD. \cite{markowitz03a} derive a slope at the lowest
frequencies $\alpha_{L}$ of $0.05^{+0.55}_{-2.05}$, a break
frequency, $\nu_{L} =1.59^{+4.73}_{-0.95} \times 10^{-6}$Hz and an
intermediate slope above the break of $\alpha_{I} =
1.20^{+0.25}_{-0.35}$. The earlier results of
\cite{pounds01} are consistent with these values.
We begin by refitting these data using our standard
simulation-based modelling method \cite[e.g. see][]{uttley02}. We
fit a simple bending power law model \cite[e.g. see][]{mch04,mch05}.
We bin the
long time-scale \xte\ light curve into 4-d bins, and the short
time-scale \xte\ light curve into 3-h bins. The resultant best-fitting
values are similar to those of \cite{markowitz03a} and the errors in
the fit parameters are similarly large.
As our value of $\alpha_{L}$ is quite
consistent with zero and as, in both their hard and very high
states, GBHs have $\alpha_{L}=0$, we fix $\alpha_{L}=0$ in
subsequent power law fits.

\subsection{Doubly-bending power law fit}

We then fit the combined \xte, \asca\ and \xmm\ data sets with a doubly
bending power law model, including a second, higher frequency break
at $\nu_{H}$ and slope $\alpha_{H}$ at the highest frequencies. We
include the \asca\ 2--10 keV SIS data in 512s bins and the
\xmm\ PN data in 48s bins in the 2--8.8 keV band to match the
average energy of the \asca\ band.
The PN bin value was chosen so that
the resultant light curve has $2^{11}$ points so that we can use the
fast Fourier transform, rather than the slower direct Fourier
transform, for that part of the simulation, without loss of accuracy.

\begin{figure}
\psfig{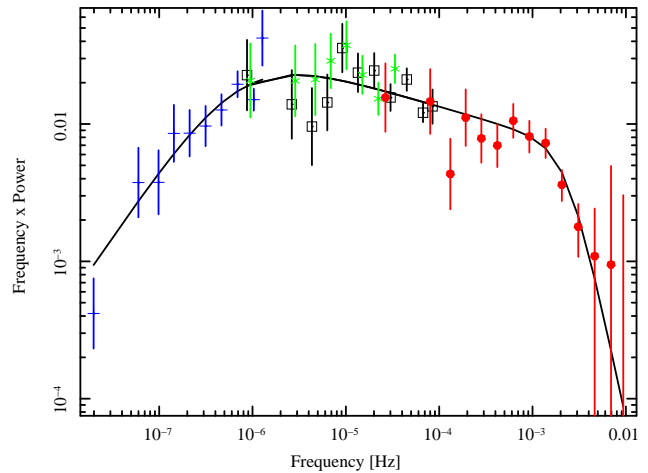}
\caption{PSD of \ark\ from \xte, \asca\ and \xmm\ data in the 2--10
  keV \xte\
energy band. The
solid black line shows the best-fitting
double-broken power law model. The symbols with error bars represent the data,
unfolded from the distortions of the observational sampling pattern.
The crosses (blue) are from the \xte\ 4-d sampling lightcurve, the
asterisks (green) are from the \xte\ 3-h sampling lightcurve, the open
squares (black) are from the \asca\ 2--10 keV SIS 512s-binned
lightcurve and the filled circles (red) are from the 48s-binned
\xmm\ 2--8.8 keV PN data.
}
\label{bendingpsd}
\end{figure}

We obtain a very good fit (acceptance
probability P=75\% - see Table~\ref{tab:2lor} and
Fig.~\ref{bendingpsd}) with $\alpha_{I}=1.2^{+0.2}_{-0.1}$,
$\nu_{L}=7.5^{+28.1}_{-5.5} \times 10^{-7}$Hz and
$\nu_{H}=2.4^{+2.3}_{-0.9} \times 10^{-3}$ Hz.
These break frequencies are in good agreement with the values obtained by
\cite{papadakis02} from analysis of \asca\ data.

\begin{figure}
\psfig{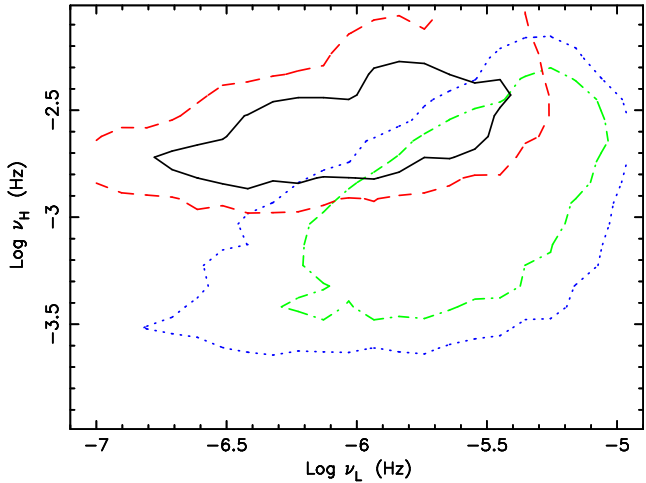}
\caption{68 \% and 90 \% confidence contours for high and low bend
  frequencies of the doubly-bending model fit to the combined
  PSD. Hard band contours (black and red respectively) are plotted in
  solid and dashed lines, soft band contours (green and blue
  respectively) in dot-dashed and dotted lines.}
\label{dbknpo_contour}
\end{figure}

In order to make a combined PSD fit of the soft X-ray band, we first
compared the low frequency variability power of the \asca\ SIS data in
the 0.6--2 and 2--10 keV bands. The resultant power spectra showed identical
amplitudes, so both energy bands can be fit together at low
frequencies without needing to rescale their PSDs. We assume that the
close similarity of the 0.6--2 and 2--10 keV lightcurves continues to
even lower frequencies, i.e. those sampled by the \xte\ long term data.
We therefore fitted
the \xte\ 2--10 keV long term data together with the SIS and \xmm\
0.6--2 keV data to cover the same frequency range as above. The fit is
again very good (P=85\% - Table~\ref{tab:2lor}) giving parameters
$\alpha_{I}=1.5^{+0.1}_{-0.4}$,
$\nu_{L}=3.6^{+5.1}_{-2.9} \times 10^{-6}$ Hz and
$\nu_{H}=1.8^{+1.8}_{-1.4} \times 10^{-3}$ Hz. The steeper
intermediate slope reflects the lower high-frequency power in the soft
band compared to the hard band. The break frequencies in both bands
are consistent within 1 $\sigma$ uncertainties, as can be seen from
the confidence contours plotted in Fig.~\ref{dbknpo_contour}

\subsection{Double Lorentzian fit to the power spectrum}

Motivated by the strong evidence for a Lorentzian component in the
\xmm\ data alone, we have fitted the combined \xte, \asca\ and
\xmm\ data with a model consisting of two Lorentzian components,

\begin{equation}
P(\nu)=A_{\rm L}\frac{Q_{\rm L}\nu_{\rm c, L}}{\nu_{\rm c,
L}^2+4Q_{\rm L}^2(\nu -\nu_{\rm c, L})^2}+A_{\rm H}\frac{Q_{\rm
H}\nu_{\rm c, H}}{\nu_{\rm c, H}^2+4Q_{\rm H}^2(\nu -\nu_{\rm c,
H})^2}
\end{equation}
where the Lorentzian peak frequency $\nu_{\rm peak}$ is related to the
centroid frequency $\nu_{\rm c}$ by $\nu_{\rm peak}= \nu_{\rm
  c}(1+1/4Q^2)^{1/2}$ and the subscripts L and H refer to the low and
high frequency components, respectively.  We fitted the combined PSD
in soft and hard energy bands independently, allowing all parameters
to vary. The low frequency component converged to very low values of
the quality factor $Q=\nu_c/\Delta \nu_{\rm FWHM}$, i.e. to very wide
Lorentzians where the Lorentizan shape is almost insensitive to the
value of $Q$. We therefore fixed this parameter at $Q=0.01$ (with 90\%
and 99\% upper confidence limits of $Q=0.2$ and 0.5 respectively) for both
energy bands to enable a more detailed parameter grid search for the
other parameters, and hence to obtain better constraints on these
parameters. The best-fitting values are listed in
Table~\ref{tab:2lor}.  The Lorentzian peak frequencies are similar to,
but not exactly the same as, the corresponding break frequencies, with
the Lorentzian frequencies lying closer to the centre of the
band-limited power.

\begin{figure}
\psfig{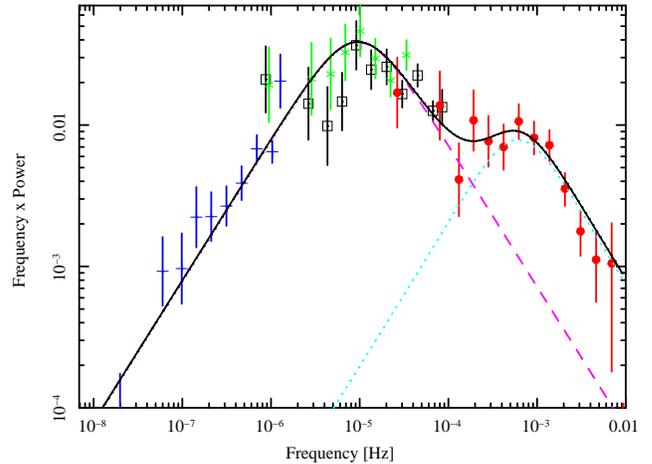}
\caption{\label{hardpsd} PSD of \ark\ from \xte, \asca\ and \xmm\ data
  in the 2--10 keV \xte\ energy band. 
 The dashed (red) line shows the low frequency Lorentzian,
the dotted (blue) line shows the high frequency Lorentzian and
the solid black line shows the
  best-fitting combined double Lorentzian model. The symbols with error bars,
  which are the same as in Fig.~\ref{bendingpsd}, represent the data
  unfolded from the distortions of the observational sampling
  pattern.}
\end{figure}

\begin{figure}
\psfig{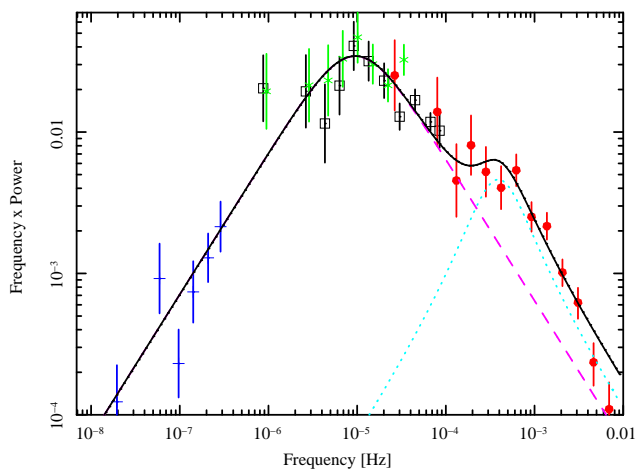}
\caption{PSD of \ark\ from \xte\ in the 2--10 keV band and \asca\ and
  \xmm\ data in the 0.6--2 keV energy band. The solid line shows the
  best-fitting double Lorentzian model and symbols with errorbars,
  which are the same as in Fig.~\ref{bendingpsd}, represent the
data, unfolded from the distortions of the observational sampling
pattern.
\label{softpsd}}
\end{figure}

\begin{figure}
\psfig{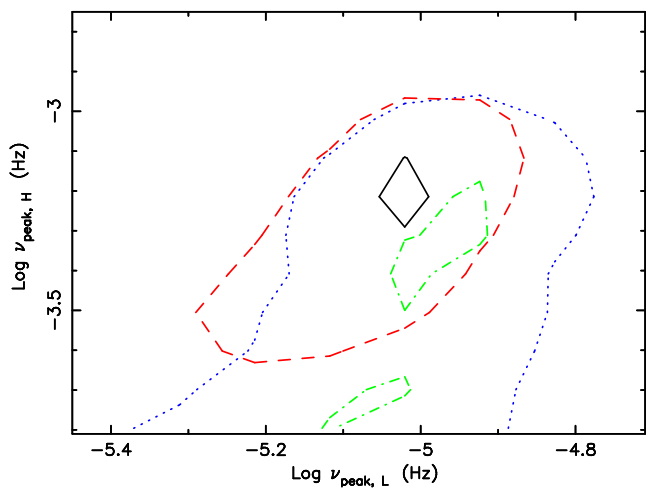}
\caption{68 \% and 90 \% confidence contours for high and low peak
  frequencies of the two-Lorentzian model fit to the combined
  PSD. Hard band contours are plotted in solid black and dashed red
  lines, soft band contours in dot-dashed green and dotted blue
  lines.}
\label{2lor_contour}
\end{figure}

Figures~\ref{hardpsd} and \ref{softpsd} show the best-fitting
two-Lorentzian models to the hard and soft band respectively. In this
interpretation, the increase in variability power at high frequencies
in the hard band is entirely due to the greater strength of the high
frequency Lorentzian in this band. The peak frequencies in both energy
bands are consistent within the 90\% confidence limits (see contour
plot in Fig.\ref{2lor_contour}).

Note that Figs.~\ref{hardpsd} and \ref{softpsd} show the PSD unfolded
through the distortions produced by sampling effects. These
distortions also depend on the underlying PSD shape so the resultant
unfolded PSD depends on the model being fitted.
If the two-Lorentzian model is the correct underlying PSD, the strong
power component around $10^{-5}$Hz will lead to spurious apparent
power at lower frequencies in a PSD which has not been unfolded from
the distortions, and where the low frequency part of the PSD has been
derived from observations which do not sample the $10^{-5}$Hz
region. This effect is known as aliasing \citep[e.g. see][for more
  details]{uttley02} and here particularly affects the long timescale
\xte\ data (blue crosses in Figs~\ref{bendingpsd},~\ref{hardpsd} and
\ref{softpsd}).  When we unfold the PSD we
remove this spurious low frequency power. Thus the lowest
frequency part of the PSDs in Figs.~\ref{hardpsd} and \ref{softpsd}
lies below the same part in the bending power law model
(Fig.\ref{bendingpsd}), where the model did not contain so much power
at higher frequencies.  Therefore the low break frequency in
Fig.~\ref{bendingpsd} is lower than in Figs.~\ref{hardpsd} and
\ref{softpsd}.

\begin{figure}
\psfig{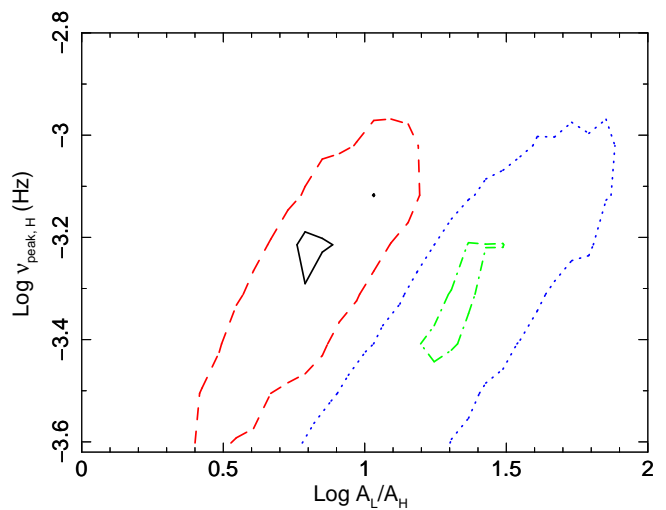}
\caption{
68\% and 90\% contours for high-frequency Lorentzian peak frequency
versus the ratio
of low and high-frequency Lorentzian normalisations, $A_{L}/A_{H}$.
Hard band contours
are plotted in solid black and dashed red lines, soft band contours in
dot-dashed green and dotted blue lines.
}
\label{relnormcon}
\end{figure}

Since the normalisations of low and high-frequency Lorentzians are
highly correlated with each other and with the Lorentzian frequency,
we only treat the ratio of the low and high-frequency Lorentzian
normalisations, $A_{L}/A_{H}$, as a free parameter and do not fit the
high and low-frequency normalisations separately.  Although the
1-dimensional errors quoted in Table~3 seem to indicate that the
normalisation ratios for the hard and soft band overlap, we show in
Fig.~\ref{relnormcon} that the 90\% confidence contours of $\nu_{\rm
  peak, H}$ versus $A_{L}/A_{H}$ do not overlap.  Therefore, for the
same $\nu_{\rm peak, H}$, the high-frequency Lorentzian
is always relatively stronger, compared to the low-frequency
Lorentzian, in the hard band than in the soft band.  In fact, the low-frequency
Lorentzian normalisation is similar in both bands and relatively
well-constrained.  For the best fitting model parameters, the values
of fractional rms in the low frequency Lorentzian are $35\pm1$\% and
$33\pm1$\% in hard and soft bands respectively, and in the high
frequency Lorentzian the values are $15\pm1$\% and $10\pm1$\% in hard
and soft bands.  Thus the low frequency Lorentzian rms values are
consistent with being the same, while the high frequency Lorenztian is
weaker in the soft band, which explains the different hard and soft
band PSD shapes at high frequencies.

\begin{table*}
\centering
\caption {\bf Fit to the PSD using combined \xmm, \asca\ and \xte\ data}
\begin{tabular}{ccccccc}
{\bf Doubly-bending power law}\\
Energy Band &$\nu_{\rm L}$[Hz]&$\nu_{\rm
H}$[Hz]&$\alpha_L$&$\alpha_I$&$\alpha_H$ &fit probability\\
\hline
0.6--2.0&3.6$^{+5.1}_{-2.9}\times 10^{-6}$ & 1.8$^{+1.8}_{-1.4}\times
10^{-3}$ &0.0 &1.5$^{+0.1}_{-0.4}$&3.4$^{+*}_{-1}$&0.85\\
2.0--8.8& 7.5$^{+28.1}_{-5.5}\times 10^{-7}$ &
2.4$^{+2.3}_{-0.9}\times 10^{-3}$
&0.0&1.2$^{+0.2}_{-0.1}$&4.2$^{+*}_{-1.8}$&0.75\\
\hline\\
{\bf Two-Lorentzian model}\\&\multicolumn{2}{c}{low-$\nu$
Lorentzian}&\multicolumn{2}{c}{high-$\nu$ Lorentzian}\\
Energy Band &$\nu_{\rm peak,L}$ [Hz]&$Q_L$ &$\nu_{\rm
peak,H}$[Hz]&$Q_H$ &$A_{\rm L}/A_{\rm H}$&fit probability\\
\hline
0.6--2.0&$9.5^{+2.4}_{-1.9}\times 10^{-6} $&$0.01$&$ 3.9^{+2.2}_{-2.3}
\times 10^{-4}$&$0.5^{+0.5}_{-0.38}$&$17.6^{+290.4}_{-11.9}$&0.37\\
2.0--8.8&  $9.5^{+2.0}_{-1.5}\times 10^{-6}$
&0.01&$6.1^{+1.5}_{-1.3}\times 10^{-4}$
&$0.125^{+0.38}_{-*}$&$6.15^{+4.7}_{-1.0}$&0.40\\
\hline
\end{tabular}

* The errors are large and are not very well defined.\\
All errors are 68\% confidence ($1 \sigma$).
\label{tab:2lor}
\end{table*}

\section{Implication of Lorentzian PSD shape on Time Lags between
energy bands}
\label{s_lags}

In Galactic black hole systems, lags are observed between light
curves in different energy bands, with the higher energy variations
generally lagging the lower
energy variations (referred to here as a `positive' lag).  If the
light curves in the different bands are separated into different
Fourier components, it is generally found that the time lags between bands
vary as a function of the Fourier frequency, i.e. time-scale, of the
component, with larger lags being associated with longer
time-scales.  (See \cite{nowak99} for a full discussion of the
relevant analysis techniques).  Such lags have now been measured in
some AGN \cite[e.g][]{papadakis01,mch04}, including in
\ark\ \citep{arevalo06}.

\cite{arevalo06} show that, for \ark , the positive time lag between
the 0.5--2 and 2--8.8 keV \xmm\ bands is approximately 800 s at low Fourier
frequencies and $\sim 0$s at high frequencies, with a sharp drop in
time lag values at a frequency of $\sim 10^{-4}$ Hz. This stepped
appearance of the lag spectrum resembles that seen in Cyg~X-1 in the
low/hard and intermediate states, where the steps in the lag spectrum
correspond to the transition frequencies between peaks in the
corresponding PSDs \citep{nowak99,nowak00}. The correspondence between
lag spectrum steps and PSD peaks suggests that each constant-lag value
is associated with a given variability component, seen as a Lorentzian
component in the PSD. Therefore, the shape of the lag spectrum in
\ark\ suggests that the two-component interpretation of the PSD is
correct.

\subsection{Fitting the two-Lorentzian model jointly to the PSD and lag spectra}

To test the above hypothesis we directly fitted the lag spectrum with a
model corresponding to the lags expected from two Lorentzian
variability components, each with a single distinct time lag, which
is constant for all Fourier frequencies.  The observed lag at any
particular frequency is then a
function of the overlap between the two Lorentzians, and must be
calculated in the complex Fourier domain, by evaluating the
cross-spectrum as a function of Fourier frequency, $\nu$.  The
cross-spectrum is given by $C(\nu)=S(\nu)H^{*}(\nu)$, where $H(\nu)$ and
$S(\nu)$ are the Fourier transforms of the hard and soft band light
curves and the asterisk denotes the complex conjugate.
$S(\nu)=S_l(\nu)+S_h(\nu)$ where the subscripts $l$ and $h$ denote the low
and high-frequency Lorentzians respectively, and similarly for the
hard band, $H(\nu)=H_l(\nu)+H_h(\nu)$.  For the low and high frequency
Lorentzian components, the soft band light curve is correlated with
the hard band light curve except with a phase lag $\phi_l(\nu)$,
$\phi_h(\nu)$ respectively.  We determine the phase lag from the fixed
low and high-frequency Lorentzian {\it time} lags $\tau_l$ and
$\tau_h$, by using $\phi(\nu)=2\pi \nu \tau$.  To avoid clutter, we now
drop the frequency-dependence of the various parameters and write the
real part of the cross-spectrum as follows:
\begin{eqnarray} Re(C)&=&Re(S)\times Re(H)+Im(S)\times Im(H)\\
&=&[Re(S_l)+Re(S_h)][Re(H_l)+Re(H_h)]\nonumber\\
  &+&[Im(S_l)+Im(S_h)][Im(H_l)+Im(H_h)]\\ &=&[|S_l|
    \cos(\phi_0+\phi_l)+|S_h|
    \cos(\phi_1+\phi_h)]\nonumber\\ &\times&[|H_l| \cos(\phi_0)+|H_h|
    \cos(\phi_1)]\nonumber\\ &+&[|S_l| \sin(\phi_0+\phi_l)+|S_h|
    \sin(\phi_1+\phi_h)]\nonumber\\ &\times&[|H_l| \sin(\phi_0)+|H_h|
    \sin(\phi_1)], \end{eqnarray} where $Re$ and $Im$ denote real and
imaginary components.  $\phi_0$ and $\phi_1$ are simply random
phase values for the low and high frequency Lorentzians respectively,
measured for each realisation of the light curves, and will be
different for different realisations. They are not related to the
phase lags of interest between bands, i.e. $\phi_l$ and $\phi_h$ for
the low and high frequency Lorentzians.  For a noise process, such as
the broad Lorentzians observed here, $\phi_0$ and $\phi_1$, in
independent Fourier frequencies, are independently and identically
distributed over a uniform distribution between 0 and
$2\pi$.  The coefficients
of the cross terms $|S_l||H_h|$ and $ |H_l||S_h|$ will average to 0
when taking the average value of the cross spectrum for many
realizations of the light curves, or many frequencies within a
frequency bin, and the average value of $Re(C)$ is then

\begin{equation}
\label{eqn:rec}
\langle Re(C)\rangle=|S_l||H_l|\langle\cos{\phi_l}\rangle+|S_h||H_h|\langle\cos{\phi_h}\rangle
\end{equation}
Similarly we can show that $Im(C)$ is given by
\begin{equation}
\label{eqn:imc}
\langle Im(C)\rangle=|S_l||H_l|\langle\sin{\phi_l}\rangle+|S_h||H_h|\langle\sin{\phi_h}\rangle
\end{equation}

\begin{figure} 
\psfig{figure=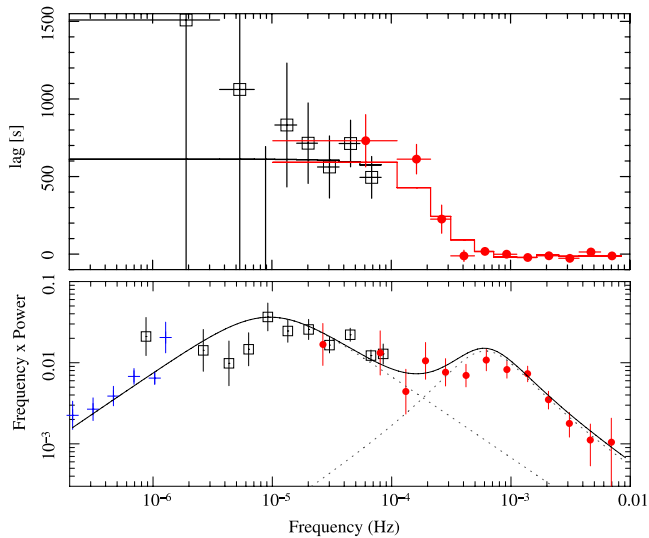,width=8.5cm}
\caption{Top panel:Lag of soft (0.5--2 keV) band to hard (2--8.8 keV)
  band, as a function of frequency, calculated using \asca\ and
  \xmm\ data. The solid line represents the lag spectrum expected for
  a two-Lorentzian PSD model with the parameters obtained by fitting
  the PSD, forcing the Lorentizan components of the hard and soft bands to
  be equal. Each Lorentzian is assumed to have a constant time lag,
  whose values were obtained by fitting the lag spectrum, resulting in
  $638\pm 54$s (i.e. hard band lags) and $-11.0 \pm 4.3$s (hard band
  leads) for the low and high frequency Lorentzians, respectively.
  Bottom panel:hard band PSD as shown in Fig~\ref{hardpsd}, with
  individual Lorentzian components overplotted. Note how the lags (top
  panel) are approximately constant in the frequency range when any
  one Lorentzian component dominates, and how the lags rapidly change
  in the frequency range where the dominance of individual Lorentzian
  components crosses over.}
\label{fitlags}
\end{figure}

$|S_l|$, $|H_l|$, $|S_h|$ and $|H_h|$  correspond to the square root of the
power of those Lorentzians in each energy band, at the frequency bin
to be evaluated.  The
observed phase lag is simply the argument of the cross-spectrum, so
the observed frequency-dependent time lag, $\tau(\nu)_{obs}$ can be
evaluated from equations~\ref{eqn:rec} and \ref{eqn:imc} using:

\begin{equation}
\tau(\nu)_{obs}=\frac{1}{2\pi \nu}\tan^{-1} \frac{Im(C(\nu))}{Re(C(\nu))}.
\end{equation}
which can be fitted directly to the observed lag spectrum using the
best-fitting Lorentzian PSD model obtained in Section 5.2 and
letting the low and high-frequency time lags be free parameters.
This simple model
reproduces well the shape of the observed lag spectrum, shown in
Fig. \ref{fitlags}, with a reduced $\chi^2=1.35$ for 17 d.o.f (fit
probability=0.15). The best-fitting values are $638\pm 54$ s for the
low-frequency lag and $-11.0 \pm 4.3$ s for the high-frequency lag.

The low frequency components fitted to the PSD data in both energy
bands are identical, but the high frequency components had different
$Q$ and peak frequency values, although they agree at the 1$\sigma$
level. Forcing the high frequency components to be equal in both
energy bands, using $Q_h=0.5$ and $\nu_{\rm peak,h}=6.1 \times
10^{-4}$~Hz (both within 1$\sigma$ of the best-fitting values),
produces a better fit to the lag spectrum, with reduced $\chi^2=1.13$
for 17 d.o.f. (fit probability = 0.32), and lags consistent with the
previous fit. The consistency between the lag spectrum and PSD
significantly strengthens the interpretation of the PSD of \ark\ as
the summation of Lorentzian components, and hence as a very high state
system.

\subsection{Coherence}
 The variability in the soft and hard bands is relatively
 coherent. Measuring the degree of linear correlation between energy
 bands as a function of Fourier frequency shows that the light curves
 remain coherent (coherence, $\gamma^{2}, \sim 0.8$ where
$\gamma^{2}=0$ for uncorrelated
 light curves and $\gamma^{2}=1$ for perfectly correlated light curves) up
 to frequencies of at least $10^{-3}$ Hz (see figure 3 in \citet{arevalo06}).
 The coherence is consistent with the two-component interpretation,
 even if the variability patterns of the low and high frequency
 Lorentzians are uncorrelated. For example, if one energy band is fully
 modulated by one of the Lorentzians, while at the same time-scales the
 other energy band responds equally to both Lorentzians, the coherence
 would drop to a value of $\gamma^{2}= 0.5$. This extreme case never
occurs in the present data, so the coherence can remain over 50 \%
even if the Lorentzians are uncorrelated.
 The relative contribution of the Lorentzians to each energy bands is
most different at frequencies $\sim 2 \times 10^{-3}$ Hz, where both
Lorentzians have
 similar power in the hard band (Figs.\ref{hardpsd} and \ref{softpsd}),
 but the high-frequency Lorentzian is 10 times stronger than the low
 frequency component in the soft band. In this case, the coherence is
 expected to drop to $\sim 0.75$, as measured by \citet{arevalo06}
 (their figure 3).

It is possible that the low-frequency variability component that we
fit to the PSD as a very broad Lorentzian actually contains more
components. The available data, however, are not good enough to
disentagle extra components in the low frequency PSD. Unfortunately
the lag spectrum cannot help in this case as the long term monitoring
was performed only by \xte , which has a hard energy response.
We are therefore unable to compute a lag spectrum between hard and soft bands
as we have been able to on intermediate and short time-scales.

\section{Discussion}
\label{s_conclusions}

\subsection{Summary of PSD and lag fits}

We have fitted the PSD of \ark\ over the $10^{-8}-10^{-2}$ Hz frequency
range, combining long term archival \xte\ monitoring data, a
month-long observation by \asca\ covering intermediate frequencies and
a new 100 ks observation by \xmm\ which covers the shortest
time-scales. We confirm the existence of two breaks in the PSD, which
make this the only AGN so far observed to have band-limited
variability.

The PSD can be fitted either with a double-bending power law,
where the variability power drops at low and high frequencies, or
with a combination of two distinct variability components, peaking
close to the low and high frequency breaks, respectively.  Using
two Lorentzian-shaped components, we obtain a good fit in both the
soft 0.5--2 and hard 2--10 keV bands, with self-consistent Lorentzian
parameters between the energy bands. We note that, when fitting
the upper break in the PSD, the frequency of the fitted Lorentzian
is a factor of 4 lower than the fitted bend frequency as, in any
band-limited PSD, the Lorentzian fits will tend to move towards
the centre of the band. This difference should be considered when
combining values of PSD break timescales derived from different
fitting methods. In the Lorentzian interpretation of the PSD, the
increase in variability power at high frequencies in the hard band
is explained by a strengthening of the high frequency Lorentzian,
as is seen in GBHs \citep[e.g.][]{pott03}, while the low frequency
Lorentzian remains approximately at the same level.

The two-component interpretation is strongly supported by the lag
spectrum calculated between these energy bands using \asca\ and
\xmm\ data (Ar\'evalo et al. 2006).  The lag spectrum has a
step-like shape, where the time lags are approximately constant at
low frequencies and drop sharply to a different constant value at
around $10^{-4}$ Hz. The drop in the lag spectrum occurs exactly
at the frequency expected if each Lorentzian component has a
single time lag value associated with it. This characteristic lag
spectral shape therefore provides strong evidence that the
different Lorentzian components in the PSD arise from distinct
physical regions, e.g. different radii in the accretion disc.

\subsection{Scaling Relationships}

The lag spectrum of Ark~564 has a very similar shape to that of Cyg X-1
in its intermediate state \cite[see figure 8 of][]{arevalo06},
although the frequencies are, of course, very different.  However our
recent 3D fit \citep{mch06} shows that PSD break frequencies scale not
only inversely with black hole mass, $M$, but also approximately with
$\dot{m}_{E}$. If we apply the same scaling to the lag frequencies,
then Ark~564 and Cyg X-1 in its intermediate state coincide almost
exactly.

Ark~564 was included in our 3D fit, using the value
of $T_{B}$ given by \cite{papadakis02}, and lay close to the best-fitting 
plane.  The better measured value of $T_{B}$ which we derive here, on
the assumption of a bending power law, is identical to the value of
$T_{B}$ given by \cite{papadakis02} thus Ark~564 remains consistent
with our previous scaling relationship.

All previous studies of the scaling between AGN and GBHs have
concentrated on the high frequency part of the PSDs as the (scaled)
low frequency part of most AGN PSDs is not well measured. On the basis
of just one AGN where we are able to fit a low frequency Lorentzian
component we are therefore extremely cautious in any comparison, and
we also note that intermediate state GBH PSDs show many and varied
shapes where the separation between high and low frequency Lorentzians
is not constant \citep[e.g.][]{axelsson05,axelsson06,belloni05}.
However the separation of approximately 1.5 decades in frequency
between the two Lorentzians seen here is not wildly different from
typical separations seen in intermediate state GBHs. Thus some sort of
mass (and probably also accretion rate) related scaling probably
applies to the lower, as well as to the upper, PSD frequencies.

\subsection{Radio luminosity and `state'}

The PSD and lag spectrum indicate a VHS/intermediate state, rather
than a soft state, with a hard state being ruled out because of the
very high accretion rate. Is this state classification consistent with
other state measures, e.g. radio luminosity?  VHS/intermediate state GBHs
can be quite luminous, although transient, radio sources \cite[e.g.
  GRS1915+105,][]{millerjones05} whilst soft state GBHs are undetected
in deep radio observations. 

Ark~564 \citep[black hole mass $M \sim 2.6 \times 10^{6}
  M_\odot$,][]{botte04} is considerably more radio luminous than AGN
with well defined soft-state PSDs and similar black hole masses, e.g.
NGC4051 \citep[$1.9 \times 10^{6} M_\odot$,][]{peterson04} and
MCG-6-30-15 \citep[$4.5 \times 10^{6} M_\odot$,][]{mch05}. Both
NGC4051 \citep[\mch et al. in preparation; see also][]{christopoulou97}
and \ark \citep{lal04,schmitt01} contain flat spectrum VLBI cores
for which $L_{Rad-Ark 564} \sim 600 \times L_{Rad-NGC4051}$
(assuming redshift-derived distances).
These observations are certainly consistent with a VHS/intermediate
state interpretation for Ark~564 although the present lack of radio
detection of soft state GBHs prevents a more quantitative comparison.

If the same disc-jet coupling applies in NGC4051 and Ark~564 and the
radio luminosity of the jet scales as $\dot{m}^{1.4}$ \citep[absolute,
  not Eddington, units;][]{blandford79}, then we would expect
$L_{Rad-Ark 564} \sim 100 \times L_{Rad-NGC4051}$ \citep[deriving
  $\dot{m}$ from][]{uttleymch05}. Thus it is not yet clear whether the
higher radio luminosity of Ark~564 relative to NGC4051 is purely due
to a higher accretion rate or to some difference in jet structure.

\subsection{Physical origin of the high frequency Lorentzian timescale}

If we take the Lorentzian peak frequency ($\sim 6\times 10^{-4}$
Hz, i.e. 1666s, in the 2--8.8 keV band) as being the best
measurement of the underlying physical process responsible for the
sharp drop in power, we can consider what that timescale might
correspond to. The dynamical time-scale around the Ark~564 black
hole ($2.6 \times 10^{6} M_\odot$) is considerably shorter than
the Lorentzian peak time-scale, unless we assume that the
time-scale arises far (few tens of $R_g$) from the black hole.
There is evidence from the broad iron X-ray flourescence lines
\citep{fabian05} that the accretion disc often reaches down to
$<2R_g$, close to the innermost stable orbit of a maximally
spinning black hole black hole. If the Lorentzian peak time-scale
arises from the inner edge of such a spinning black hole disc, it
may well correspond to the viscous time-scale (assuming a
viscosity parameter, $\alpha=0.1$). In that case the ratio of disc
scale height to radius, $H/R$, = 1.2, which is reasonable given
the very high accretion rate.

\subsection{Combined interpretation of PSD and lag spectrum}

A possible interpretation of both the PSD and lag spectrum is that the
variability originates mainly in two distinct regions.  Note that by
`variability' we mean the source of the variations, not the source of
the X-rays. The variability may be related to accretion rate
fluctuations in the accretion disc. It is reasonable to assume that
variability time-scales are larger at larger radii
\cite[e.g.][]{lyubarskii97}.  Once produced, variations would then
propagate inwards and, on reaching the X-ray emitting region, would
modulate the emission. It is also reasonable to assume that the
emitted energy spectrum becomes harder towards smaller radii so, for
example, the bulk of the 0.5--2 keV emission will be emitted from a
larger radius than the bulk of the 2--10 keV emission. If the region
responsible for most of the variability at low frequencies lies
outside all of the X-ray emitting region, then the time lag between the
soft and hard bands may represent the time taken for the fluctuations
to propagate from the `soft' to `hard' emission region (e.g. see
\citealt{kotov01}, \citealt{arevalouttley06}).  If all of the
frequencies in the low frequency variability component are produced in
the same region, for example a narrow annulus of the disc,
then all of those frequencies will have the same soft-hard lag.

If the source of the higher frequency variations lies within the X-ray
emission region, e.g. between the soft and hard emission regions, then
the high frequency variations will affect only the harder emission
region and the resultant X-ray variations will be stronger at higher
energies, as we observe. In this scenario the predominantly softer,
outer emission region, which is subject only to the slower variations,
may form part, or possibly all, of the so-called `soft excess',
i.e. the excess in the energy spectrum below $\sim 2$keV above an
extrapolation of a power law to low energies. This softer region,
although lying outside the harder emitting region is, however,
probably still quite close (ie within $\sim10 R_{G}$ rather than
$\sim100 R_{G}$) to the black hole.  This scenario is consistent with
the observations of \cite{turner01} and \cite{brinkmann07} of Ark
564. \cite{turner01} show that, on $\sim1$ day time-scales, the soft
excess emission from Ark~564 varies at least as much as the power law
emission. \cite{brinkmann07} show, by fitting a combination of
bremsstrahlung (not necessarily a physical bremsstrahlung component
but simply to fit the `soft excess') and power law components to 500s
sections of the \xmm\ data discussed here, that the `bremsstrahlung'
component at low energies (0.3--1 keV) is less variable than the
power law component in the same band, and that the power law component
at higher energies (3--10 keV) is more variable still.  However a full
understanding of how the soft excess is related to the rest of the
X-ray emission, e.g. whether it might actually be caused by absorption
\citep{gierlinski06} requires detailed frequency-resolved spectroscopy
\citep[e.g. see][]{papadakis07b,revnivtsev01} for which the present
data are insufficient.

We finally comment on the negative phase lag, $-11.0\pm 4.3$s, at
the highest observable frequencies.  As the high frequency X-ray
variations at lower energies will also be produced at small radii,
e.g. as the low energy tail of the harder emission region, there
should be very little lag between the soft and hard emission, as
we see. However the lag is negative, i.e. the hard band leads, in
the higher frequency component. This observation suggests that the
high-frequency variability in the soft band might arise from
reflection/reprocessing within $\sim1R_g$ of the illuminating hard
X-ray source, implying the existence of cold gas close to the
central X-ray source.

\section*{Acknowledgements}

This work is based on observations with \xmm, an ESA science mission
with instruments and contributions directly funded by ESA Member
States and the USA (NASA). IMcH and PA acknowledge support from STFC under
rolling grant PP/D001013/1 and PU acknowledges support from an STFC
Advanced Fellowship. We acknowledge redshift data from the NASA/IPAC
Extragalactic Database.

\label{lastpage}
\end{document}